\documentstyle[12pt,aps]{revtex}
\begin{document}
\baselineskip 0.6cm
\def\SNG{{\em Physical Review Style and Notation Guide}}
\def\LUG {{\em \LaTeX{} User's Guide \& Reference Manual}}
\def\btt#1{{\tt$\backslash$\string#1}}%
\def\REVTeX{REV\TeX}
\def\AmS{{\protect\the\textfont2
        A\kern-.1667em\lower.5ex\hbox{M}\kern-.125emS}}
\def\AmSLaTeX{\AmS-\LaTeX}
\def\BibTeX{\rm B{\sc ib}\TeX}


\title{ Local impurities in 2D Quantum Antiferromagnets}
       
\author{V.N. Kotov$^{*}$ , J. Oitmaa, O. Sushkov }

\vspace{3.6cm}
\address{ School of Physics, University of New South Wales,
 Sydney 2052, Australia}

\maketitle

\vspace{.5cm}
\begin{abstract}

\normalsize{
\baselineskip 0.6cm
 The 2D quantum Heisenberg antiferromagnet
at zero temperature with two 
kinds of locally frustrating defects -  ferromagnetic bonds
and impurity spins, is studied. An analytic approach is developed to
investigate strong defect-environment couplings. 
Results for the local ground state properties
 are compared to numerical simulations and previous
work.

}
\vspace{1.0cm}
\end{abstract}

\vspace*{13.0cm}

$^{*}$Corresponding Author: Fax: +61-2-9385-6060,\\
 E-mail: vnk@newt.phys.unsw.edu.au

\clearpage

 The two-dimensional Heisenberg antiferromagnet (HAFM) has attracted
a lot of attention recently, mainly because of its relevance to
the physics of the high-$T_{c}$ materials.
  Doping with holes, which is crucial for the superconducting
properties, leads to frustration of the spins and ultimately
to destruction of antiferromagnetic (AFM) order. While the
holes generally can hop, the extreme limit of static holes,
 acting as locally frustrating defects, is believed to
describe some of the physics \cite{Aharony,Manousakis}.

 We study the Hamiltonian:

\begin{equation}
  H = J \sum_{<i,j>}\vec{S}_{i} . \vec{S}_{j} -
     K \vec{S}_{1} . \vec{S}_{2}+ L \vec{S}_{0} . 
 (\vec{S}_{1} + \vec{S}_{2}).
\end{equation}

 The first term is the HAFM on a square lattice with nearest neighbor
interactions  
(we set $J=1$ from now on).
The second and third term describe, respectively, a 
 ferromagnetic bond (FMB) of strength $K\geq0$, connecting
the neighbor sites 1 and 2, and a quantum impurity spin $\vec{S}_{0}$, coupled
to those    two sites. All the spins are 1/2.
 Without the two perturbations, the HAFM is known
to have an ordered two-sublattice  ground state
\cite{Manousakis}.
Both additional terms lead to local frustration and thus influence
the AFM order in the neighborhood of the defect.

 First, we reconsider the case of a FMB in an AFM
environment $(L=0, K\neq0)$, previously studied in Ref.[3] in the
linear spin-wave approximation (LSWA).
 Since the perturbation  is local, it is  natural first to solve 
the problem of the two spins, connected by the bond, and then 
 take into account their interaction with the AFM environment (spin waves).
Technically, we compute   the two-particle Green's function
of the spins 1 and 2. Details can be found in Ref.[5].
The interaction Hamiltonian is,

\begin{equation}
H_{int} = \frac{1}{2}   \{ S^{+}_{1} \sum_{nn1
             }a^{+}_{i} + S^{+}_{2} \sum_{nn2} b_{j} + h.c.  \},
\end{equation}
which represents the transverse part of the interaction of the
spins 1 and 2 with their neighbors (the sums stand for the three nearest 
neighbors only, excluding the other spin, connected by the bond).
 Here $a$ and $b$ are the Holstein-Primakoff operators on 
the two sublattices \cite{Manousakis}. The full Hamiltonian is a sum of
$H_{int}$, the Ising part of the interactions in $H_{int}$  and the usual free
spin wave term. Since the defect disturbs the AFM order away from it very
weakly, one can treat the Ising terms in  mean-field, by replacing 
$S_{1}^{z}S_{i}^{z} \rightarrow S_{1}^{z}<S_{i}^{z}>$, where $i$ is any
neighbor of $1$, and similarly for the other spin. We take
$<S_{i}^{z}> = 0.303$, which is the magnetization of the clean  HAFM
in LSWA \cite{Manousakis}. 

 We have developed a perturbation theory
for $H_{int}$, keeping only the leading (one-loop) contributions
to the self-energy \cite{us}. 
For the unperturbed  case $(K=0)$, this approach gives
\mbox{ $M \equiv <S_{1}^{z}>=0.316$},
 a slightly higher value than the LSWA. Our results for different
values of $K$ are summarized in Fig.1., where the correlation
function \mbox{$C(1,2) = <\vec{S_{1}} . \vec{S_{2}}>$} is also plotted. 
 Additionally, we have performed numerical simulations, based on exact
diagonalization on small clusters, with very similar results,
which, for lack of space, are not presented on the figure.
The numerical procedure we have used differs from the one in previous
studies \cite{Jaan}.
 In  the spirit of our analytic approach,
a staggered magnetic field of magnitude $|h| = 0.303$ was
applied to the spins on the boundary of the cluster (typically
containing $N=18$ spins). An additional advantage of using such
a field is that it breaks the sublattice symmetry, thus allowing
for a non-zero  staggered magnetization.

 The behavior of the local  magnetization within our
approach is clearly quite different from the LSWA result.
 Up to $K=1$ the two curves follow each other closely, both 
predicting a slight increase of $M$ at that value (corresponding
to a missing bond).
For larger $K$ the magnetization in LSWA drops sharply,
eventually goes through $0$ and then
 diverges at $K=2$, signaling an
instability, which is related to a local triplet
formation. Of course, one expects to be below the   range of validity
of the LSWA even below $K=2$, since the number of generated
spin waves is large  and interactions between them become
important. 
  
 Our result for the
magnetization clearly shows that the range of validity of
LSWA  is limited to small values of $K$, up to $K=1$. Beyond
this point the local magnetization decreases slowly and stays
non-zero  even for large $K$.  
On the other hand, the correlations across the FMB change from
  AFM to ferromagnetic at $K \approx 2.1$. One can show that
it is the transverse (in plane) part of the correlator, which
changes sign, while the Ising part shows AFM correlations for
all $K$ \cite{us}. With increasing $K$ the transverse part increases
and  the Ising one decreases (in magnitude), but remains non-zero,
which is consistent with a finite magnetization. 
 To conclude, our approach naturally describes the local triplet
formation at $K \approx 2$; moreover, it can be used far
beyond this point.

 Next, we discuss the case $L\neq0, K=1$, corresponding to a quantum
impurity spin in an AFM background (the bond between
the spins, interacting with the impurity, is subtracted). 
 We consider the three-particle
Green's function (of the impurity spin and its neighbors), and 
apply the method described above \cite{us}. The results are summarized in
Fig.2. Again, only  our analytic results are presented;
the  numerical simulations follow closely the plotted curves.
 The magnetization is  defined in the usual
way: $M(i) = < S_{i}^{z}>$.
It is assumed that for $L=0$ the spin $ \vec{S}_{1}$ belongs to   sublattice
A (spin up), $ \vec{S}_{2}$ - to sublattice B (spin down) and
$S_{0}^{z} = 1/2$. 
 For  ferromagnetic coupling $L \leq 0$, all correlations 
become ferromagnetic for sufficiently large $|L|$. However, 
this is not accompanied by a change in the ground state, since
all the three spins have a non-zero magnetization (Fig.2.). This was
already pointed out in Ref.[4]. 
 For antiferromagnetic coupling $L \geq 0$, the spins 0 and 2
change their direction at $L \approx 2.3$ ($M(0)$ and $M(2)$ change
 sign). Naturally, the magnetization
on site 1 increases at this point, since   $\vec{S}_{1}$ is
 no longer frustrated.
 This local ground state phase transition has  already been observed
 numerically in Ref.[4] for a finite size system of N=18+1 spins.
 However, a finite jump in all correlation functions, as well as
magnetizations was reported there, while we  observe a continuous
behavior. We believe the discontinuity to be a finite size effect.

 In summary, we have  developed a technique, which allows
us to study the effects of locally frustrating perturbations             
in  2D quantum antiferromagnets for any strength 
of the coupling.  For an isolated FMB we find  that the local 
magnetization does not vanish even for strong coupling - 
a result, suggesting that the previously used LSWA leads
to qualitatively wrong behavior, and its region of
validity is limited to small coupling only. We also
study frustration, produced by a quantum impurity spin,
coupled symmetrically to the two sublattices and describe
the local ground state transition  which we find in this case
for antiferromagnetic sign of the coupling.

 This work is supported by the Australian Research Council.

\clearpage

 FIGURE CAPTIONS

\vspace{1.0cm}

 Fig.1. The magnetization $M$(solid line) and the correlation function
$C(1,2)$ (long dashed line) as a function
of the ferromagnetic bond's strength. The short dashed
line is the magnetization, calculated in LSWA \cite{Lee}. 

\vspace{.5cm}
 Fig.2. Local magnetization at the impurity and the neighbor sites
 versus interaction strength.
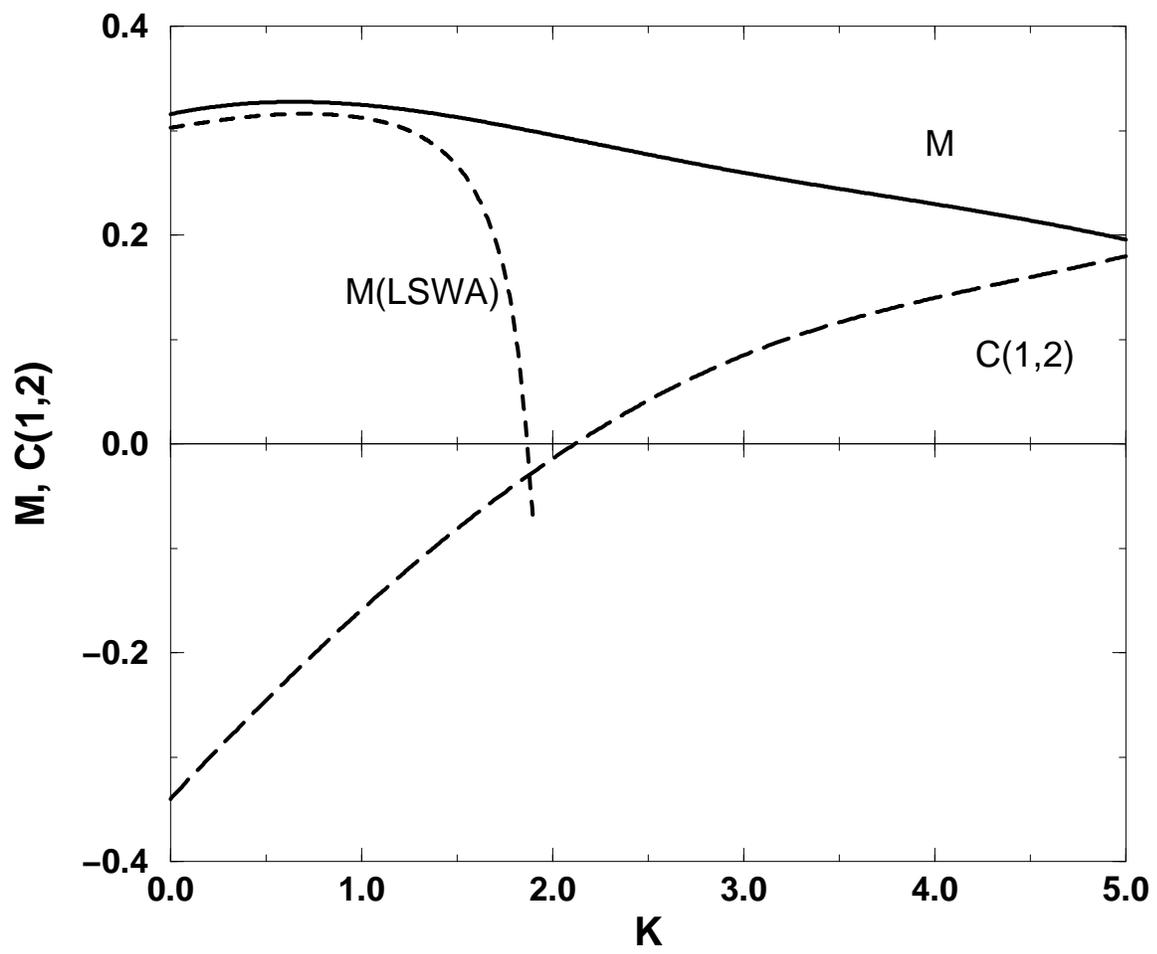
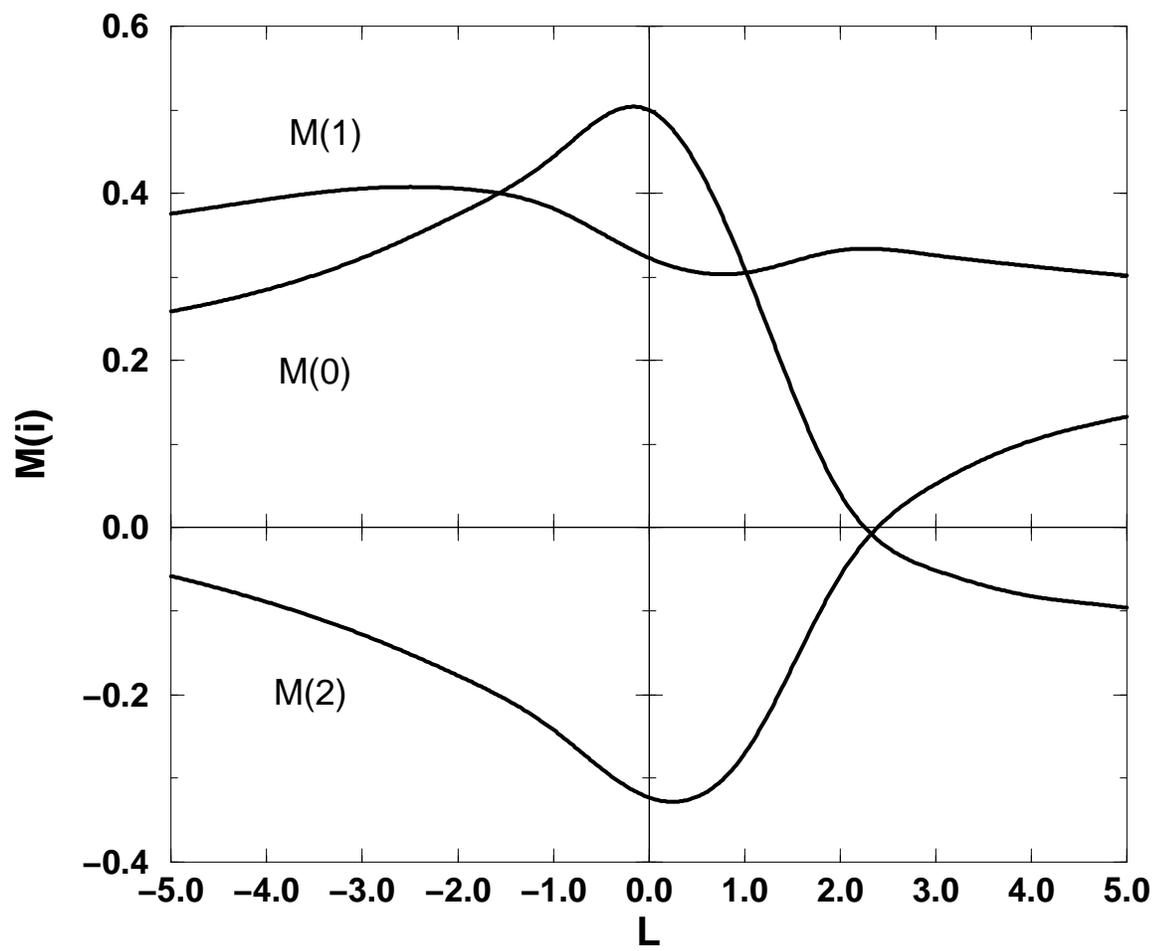
\end{document}